\documentclass[aip, pop, reprint]{revtex4-1}

\usepackage[english]{babel}
\usepackage{times}
\usepackage{epsfig,graphicx}
\usepackage[tight]{units}
\usepackage{amsmath} 
\usepackage{comment} 
\usepackage{siunitx} 

\begin{document}

\title{Electron vortex magnetic holes: a nonlinear coherent plasma structure}

\author{Christopher T. Haynes}
\email{c.t.haynes@qmul.ac.uk}
\affiliation{School of Physics and Astronomy, Queen Mary University of London, Mile End Road, London E1 4NS, UK}

\author{David Burgess}
\affiliation{School of Physics and Astronomy, Queen Mary University of London, Mile End Road, London E1 4NS, UK}
\author{Enrico Camporeale}
\affiliation{Multiscale Dynamics, Centrum Wiskunde \& Informatica (CWI), Amsterdam, Netherlands}
\author{Torbjorn Sundberg}
\affiliation{School of Physics and Astronomy, Queen Mary University of London, Mile End Road, London E1 4NS, UK}

\begin{abstract}
We report the properties of a novel type of sub-proton scale magnetic hole found in two dimensional PIC simulations of decaying turbulence with a guide field. The simulations were performed with a realistic value for ion to electron mass ratio. These structures, electron vortex magnetic holes (EVMHs), have circular cross-section. The magnetic field depression is associated with a diamagnetic
azimuthal current provided by a population of
trapped electrons in petal-like orbits. The trapped electron population provides a mean azimuthal velocity and since trapping preferentially selects high pitch angles, a perpendicular temperature anisotropy. The structures arise out of initial perturbations in the course of the turbulent evolution of the plasma, and are stable over at least 100 electron gyroperiods. We have verified the model for the EVMH by carrying out test particle and PIC simulations of isolated structures in a uniform plasma. It is found that (quasi-)stable structures can be formed provided that there is some initial perpendicular temperature anisotropy at the structure location. The properties of these structures (scale size, trapped population etc.) are able to explain the observed properties of magnetic holes in the terrestrial plasma sheet. EVMHs may also contribute to turbulence properties, such as intermittency, at short scale lengths in other astrophysical plasmas.    
  
\end{abstract}

\maketitle

\section{Introduction}
The term ``magnetic hole'' (MH) was first used by Turner et al.\cite{Turner:1977} who identified large dips in the magnetic field amplitude in the solar wind, in an otherwise undisturbed background. For a subclass of such events, sometimes described as ``linear,'' the magnetic field direction remains unchanged through the event. From solar wind observations, these structures have sizes of tens to hundreds of proton thermal gyroradii, and are found predominantly in regions where the conditions are marginally mirror stable\cite{Winterhalter:1994}. Similar MH structures, either as isolated events or as trains of structures, have now been observed across a wide range of plasma environments, in planetary magnetosheaths\cite{Cattaneo:1998,Genot:2009}, cometary environments \citep{Russell:1987}, the solar wind\cite{Turner:1977,Zhang:2009} and in the heliosheath\cite{Burlaga:2006}.

The association of magnetic holes with regions of flow compression and enhanced perpendicular temperature anisotropy, such as the magnetosheath downstream of planetary bow shocks, indicates an association with the linear mirror instability with threshold $\beta_\perp /\beta_\parallel > 1 + 1/\beta_\perp$, where $\beta=2\mu_0nk_BT/B^2$. The mirror wave driven by this instability is non-propagating, has wavevector highly oblique to the magnetic field direction, is predominantly linearly polarized, and for high $\beta$ has a higher growth rate than the cyclotron anisotropy instability, which is the competing low frequency instability driven by a perpendicular temperature anisotropy\cite{Gary:1993}. The observed sized of MH in the solar wind and planetary magnetosheaths corresponds to scales larger than proton thermal gyroradius, in agreement with the range of unstable wavelength found for the mirror instability\cite{Gary:1993}.

The development of nonlinear mirror structures to form magnetic holes has been the subject of several studies. In a high $\beta$ plasma the nonlinear evolution of the mirror instability can lead to the formation of both magnetic holes and peaks, depending on whether the plasma is linearly mirror stable or unstable (respectively)\cite{Genot:2009}. This appears to be tied to a bistability phenomenon, whereby nonlinear mirror structures which are magnetic holes can exist in a plasma which is below the mirror linear instability threshold. This behavior is found in both simulations\cite{Baumgartel:2003} and fluid models based on anisotropic MHD with a Landau fluid closure\cite{Passot:2006}. Observations of magnetic holes in the solar wind and magnetosheath are consistent with them being non-propagating structures, such as mirror structures, but would be incompatible with dark soliton solutions of the Hall-MHD system\cite{Baumgartel:1999}.

In the context of coherent nonlinear plasma structures, it is interesting to consider whether there can be magnetic holes at smaller, sub-proton scales. Electron phase space holes have been long studied as examples of nonlinear solitary structures, either as solutions of the unmagnetized Vlasov-Maxwell system (BGK waves)\cite{Bernstein:1957}, the Vlasov-Poisson system\cite{Schamel:2012}, or using fluid models\cite{Schamel:1979,Turikov:1984}. These structures depend for their existence on particle trapping in the electrostatic potential of the wave. A model of magnetic structure associated with electron phase space holes has been proposed by Treumann and Baumjohann\cite{Treumann:2012}, in which an azimuthal electron current is driven by the $\mathbf{E} \times \mathbf{B}$ drift, enhancing the magnetic field at the center and reducing it outside. Simulations of the formation of electron phase space holes via the nonlinear evolution of the two stream instability\cite{Wu:2012} also show evidence of magnetic perturbations associated with these structures.

It is also possible to form magnetic structure at small scales, below the proton thermal gyroradius, via the electron mirror mode/field swelling instability\cite{Marchenko:1988,Basu:1982,Basu:1984,Pokhotelov:2013}. This instability occurs if the electron temperature is greater than the ion temperature, and requires a high perpendicular electron temperature anisotropy, $T_{e\perp} \gg T_{e\|}$. In some ways this instability is analogous to the mirror instability. A fluid approach based on electron-MHD has shown that soliton waves can be formed from the electron magnetosonic wave mode\cite{Ji:2014}, providing a way to form one-dimensional electron-scale magnetic structures.

In this paper we describe magnetic holes at sub-proton scales which are seen to form in two-dimensional particle-in-cell (PIC) simulations of turbulent relaxation. The role of reconnection in the evolution of turbulence in these simulations has been discussed in a previous work\cite{camporeale:2011,Haynes:2014}. We develop a model for the coherent magnetic hole structures and demonstrate that they can exist in a uniform, stable plasma, given certain criteria for their formation. This indicates that they are a new type of coherent plasma structure, that may exist in a variety of plasma regimes.

The magnetic holes found in the simulations are non-propagating, and have a circular cross-section, with a diamagnetic azimuthal current provided by a population of trapped electrons in petal-like orbits. There is a mean azimuthal electron velocity at the edge of the magnetic hole, and for this reason we use the term ``electron vortex magnetic hole'' (EVMH) to describe them. They form out of magnetic field perturbations during the turbulent evolution of the simulation, and are associated with an increased perpendicular temperature anisotropy due to the trapped population. We show that the properties of EVMH are similar to those of sub-proton scale magnetic holes observed in the terrestrial plasma sheet\cite{Ge:2011, Sun:2012}.

This article will be split into four sections. In the following section (Section~\ref{sec:turb_sims}) we will describe the PIC simulation of turbulence, and examine the properties of the magnetic holes that form within it. In Section~\ref{sec:test_particle_sims} we examine the composition of the diamagnetic current that sustains the EVMH, by investigating the types of electron orbit possible within these structures using a test particle code, assuming static electric and magnetic fields. In Section~\ref{sec:isolated_sims} we will again use PIC simulations  to initialize these structures in a self-consistent model, in the absence of turbulence.
\section{Turbulence simulations}
\label{sec:turb_sims}

In this section we describe the formation and properties of EVMHs as found in PIC simulations of turbulence. These simulations use the code Parsek2D\cite{markidis:2009}. This code uses the implicit moment method for time advance of the electromagnetic fields, and a predictor-corrector method for the particle mover. The implicit method allows larger time steps and box sizes compared with explicit PIC methods, which are usually constrained (for numerical stability) by the condition $ \omega_{pe} \Delta t < 2$, where $\Delta t$ is the time step, and $\omega_{pe}$ is the electron plasma frequency. Also Parsek2D allows a relaxation of the Courant-Friedrichs-Lewy (CFL) condition $c\Delta t/\Delta x < 1 $, where $c$ is the speed of light and $\Delta x$ is the cell size, the time step is $\Delta t=0.05\Omega_e^{-1}$, where $\Omega_e$ is the electron gyrofrequency. The code allows cell sizes larger than a Debye length $\lambda_D$, and in these simulations $\Delta x = \Delta y \sim 17 \lambda_D$. The code is two dimensional in the $x$-$y$ plane but retains all three vector components for velocities and fields. The electron-proton plasma is initially loaded with a uniform, isotropic Maxwellian distribution. The simulation box is $200 \times 200$ cells, with periodic boundary conditions and $6400$ simulation particles per cell for each species. This large number of particles reduces the statistical particle noise so that the dynamic range in Fourier space is large enough to resolve the formation of a turbulent cascade. The simulation box is sized to resolve wave vectors ranging from $k\rho_e = 0.1$ to $k\rho_e = 10$, where $k$ is wave vector and $\rho_e$  the thermal electron gyroradius. The box length is about 1 ion inertial length, and the electron  gyro-motion is resolved with $\sim 3$ cells per electron gyroradius (based on the initial guide field strength). The ratio of ion plasma frequency to ion cyclotron frequency $\omega_{pi}/\Omega_{i}$ is about 1650, and the ion to electron mass ratio is physical with $m_i/m_e = 1836$. The ions and electrons are initialized at the same temperature $\beta_e = \beta_i = 0.5$. The simulation was run for a total period of $t = 200\Omega_e^{-1}$.

We initialize the simulation with a magnetic guide field $\mathbf{B}_0$ perpendicular to the simulation plane, in the $z$-direction, and add random long wavelength magnetic field fluctuations, using the same parameters as previous work \cite{camporeale:2011,Haynes:2014}. No spectral slope is imposed on the fluctuations, which comprise of all three components for wave vectors $k_x=2\pi m/L_x$ and $k_y=2\pi n/L_y$ for $m=-3,\ldots,3$ and $n=-3,\ldots,3$.  The initial electric field is zero, but the abrupt perturbation of the magnetic field acts to initialize the self-consistent evolution of the turbulent decay after a short period at the start of the simulation. The development of turbulence has been discussed in earlier work\cite{Haynes:2014}.

\begin{figure}
\includegraphics[scale=0.37,angle=0]{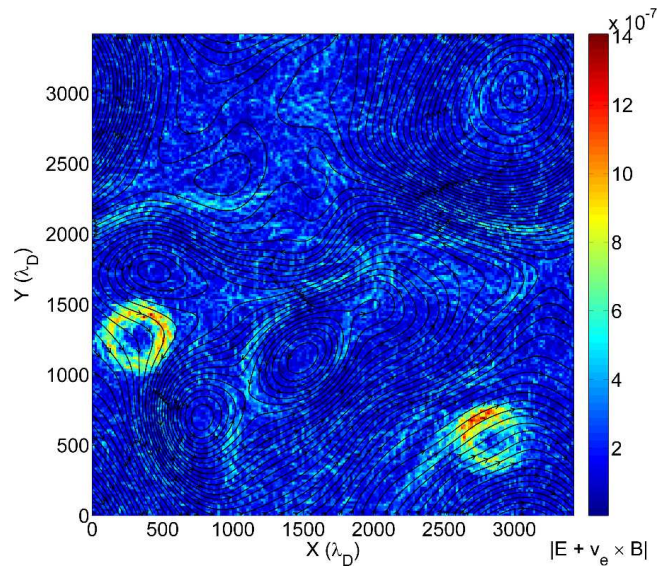}
\caption{\label{fig:EplusVxB} Magnetic field lines (black) and the parameter $|\mathbf{E}+\mathbf{v_e} \times \mathbf{B}|$ at time $t = 200\Omega_e^{-1}$.}
\end{figure}

In Fig.~\ref{fig:EplusVxB} we plot the parameter $|\mathbf{E}+\mathbf{v_e} \times \mathbf{B}|$ which in ideal, two-fluid MHD should be equal to zero. Hence, the magnetic field should be frozen into the electron fluid under these conditions, and therefore areas where $|\mathbf{E}+\mathbf{v_e} \times \mathbf{B}|$ shows large departures from zero will highlight where non-ideal (kinetic) effects are taking place. These effects could be caused by large currents, pressure gradients, wave activity, and any non-fluid-like behavior of the electrons and/or ions.

Figure~\ref{fig:EplusVxB} clearly shows, using this parameter, that two circular kinetic features have formed in the PIC simulation. Both of these features are coherent structures which remain relatively static in their appearance throughout the simulation, although they slowly drift in the $x$-$y$ plane as the simulation progresses. The structure on the left is persistent, is present from near the start of the simulation, and is likely a result of the initial field perturbation/shape in that location. The circular feature on the right hand side of Fig.~\ref{fig:EplusVxB} spontaneously forms later in the simulation, at $t \sim 100\Omega_e^{-1}$, and slowly drifts left to the position shown at $t = 200\Omega_e^{-1}$ in Fig.~\ref{fig:EplusVxB}. Although there are other features visible, in this paper we concentrate on these circular structures. 

\subsection{Plasma parameters within the structure }\label{sec:Turb_params}
\begin{figure}
\includegraphics[scale=0.22]{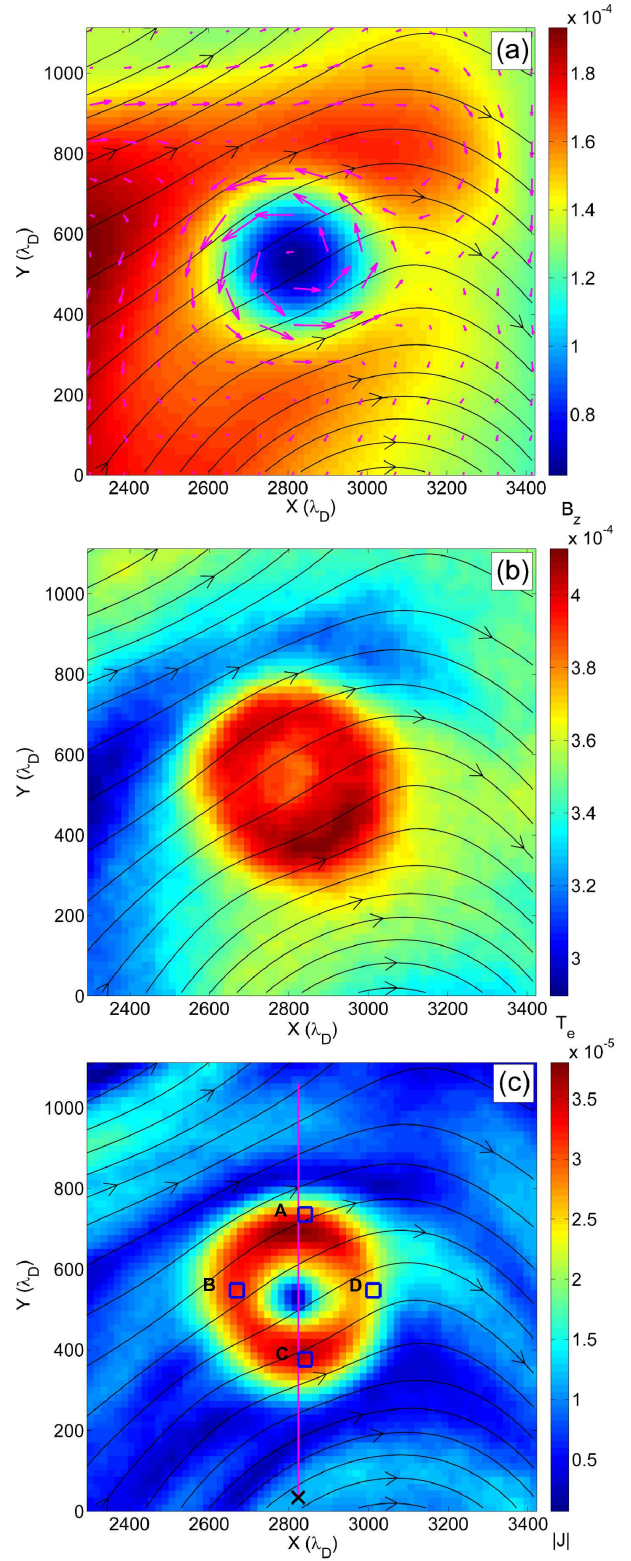}
\caption{\label{fig:2Dproperties1} Magnetic field lines (black) and (a) out-of-plane magnetic field $B_z$ with electron flow vectors (magenta), (b) total electron temperature, $T_e$ and (c) magnitude of current density, $J$. Data shown is for time $t = 200\Omega_e^{-1}$. Lengths units are in Debye lengths. The vertical line shows where cross sections of the parameters will be taken. The four marked regions are discussed in the text.}
\end{figure}

Figure~\ref{fig:2Dproperties1} shows plasma parameters in the local region of the structure on the right hand side of the simulation box as shown in Fig.~\ref{fig:EplusVxB}. In all panels in-plane magnetic field lines are plotted as black contours, with the color shading showing the value of different parameters. Figure~\ref{fig:2Dproperties1}(a) shows the out-of-plane magnetic field, $B_z$, and the direction of electron bulk velocity is shown by magenta arrows. The structure is clearly characterized by a depletion in $B_z$ of $\sim 50\%$, so we are justified in calling the structure a magnetic hole. Electron flow is in a circular anti-clockwise direction within the hole, much like a vortex, and therefore we shall refer to these structures as electron vortex magnetic holes (EVMHs). 

\begin{figure}
\includegraphics[scale=0.6,angle=0]{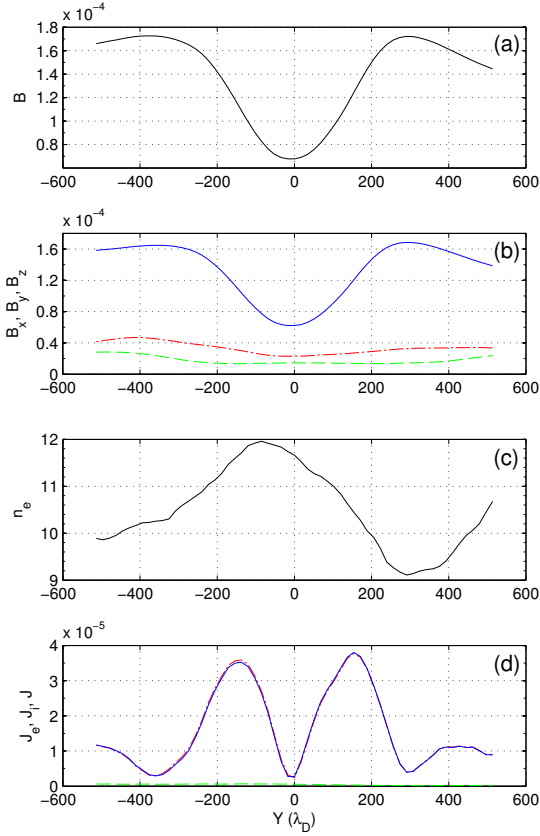}
\caption{\label{fig:xsection1} Cross section of magnetic field and current parameters through the line indicated on Fig.~\ref{fig:2Dproperties1}(c). Multiple variables (see vertical axis) are colored in the order, red (dash-dot line), green (dashed line) and blue (solid line).}
\end{figure}

\begin{figure}
\includegraphics[scale=0.7,angle=0]{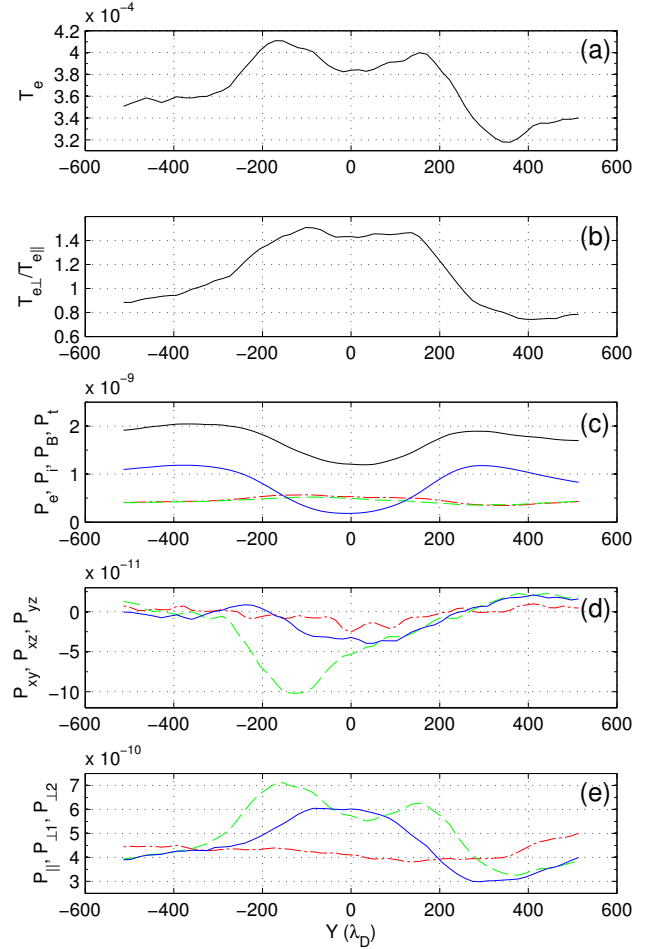}
\caption{\label{fig:xsection2} Cross section of plasma temperature and pressure parameters through the line indicated on Fig.~\ref{fig:2Dproperties1}(c). Multiple variables (see vertical axis) are colored in the order, red (dash-dot line), green (dashed line), blue (solid line) and black (solid line) respectively.}
\end{figure}

Figure~\ref{fig:2Dproperties1}(b) shows the electron temperature, $T_e$, within the EVMH and Fig.~\ref{fig:2Dproperties1}(c) shows the magnitude of current density, $J$. These panels show that the depletion in $B_z$ corresponds to a region of hot electrons. The current density in the region shows a characteristic ring shape. The vertical line in Fig.~\ref{fig:2Dproperties1}(c) was used to generate a cross section of parameters through the structure, as shown in Figs.~\ref{fig:xsection1} and \ref{fig:xsection2}. The black cross in Fig.~\ref{fig:2Dproperties1}(c) indicates the start (left edge) of the displayed cross sections in Figs. \ref{fig:xsection1} and \ref{fig:xsection2}.

The units of the horizontal axis in Figs.~\ref{fig:xsection1} and \ref{fig:xsection2} are in Debye lengths, measured from the center of the EVMH. The hole's center was determined by finding the location of minimum $B_z$ in the region. The panels in Fig.~\ref{fig:xsection1} show from (a) to (d) : $B$, ($B_x$, $B_y$, $B_z$), $n_e$, and ($J_e$, $J_i$, $J$). Number density is shown in particles \si{\per\cm\cubed}, and all other parameters are shown in normalised simulation units ($\tilde {B} \rightarrow \dfrac{B\,10\,q}{\omega_{pe}\,c\,m_i}$, and $\tilde{J} \rightarrow \dfrac{J\,100\,q}{\omega_{pe}^2\,c\,m_i}$). Where multiple parameters are plotted on a single axis, the variables are colored in the order, red (dash-dot line), green (dashed line), blue (solid line) and black (solid line).

Figure~\ref{fig:xsection1}(b) shows the shape of the depletion in $B_z$ between $\pm300\lambda_D$. This appears to be approximately sinusoidal in form (we will examine this later in Section~\ref{sec:test_particle_sims}). $B_x$ and $B_y$ are approximately constant within the structure. The depletion of $B_z$ is $\sim50\%$ at the hole's center. As the system evolves the reduction in $B_z$ becomes larger. The magnetic field outside the hole makes an angle of $\sim$\ang{15} to the $z$-axis (see Fig.~\ref{fig:xsection1}(b), at $+300\lambda_D$, $\theta \sim \arctan(1.6/0.4)$) which increases to $\sim$\ang{60} at the center of the hole ($\theta \sim \arctan(0.6/0.3)$). The electron number density increases slightly within the hole (Fig.~\ref{fig:xsection1}(c)) but does not appear to be completely anti-correlated with the field strength. The initial background simulation density was set to 10 particles per \si{\cm\cubed}. The 1-D density cut shows local fluctuations of $\pm10\%$, but the density within the hole peaks at $+20\%$. Figure~\ref{fig:xsection1}(d) shows the contribution to current density from both species, and the magnitude of the total current density, $J$. The ring-shaped current is clearly visible as two sharp peaks in the current cross section, each at approximately $\pm200\lambda_D$ from the hole's center. The radius of the hole, determined from this current density cross section, is $\sim 300 \lambda_D$. The magnitudes of the current density components show that the current is entirely generated by electron flow, as the $J_e$ line (red) and $J$ line (blue) completely overlap, and ion current (green) is negligible.

Figure~\ref{fig:xsection2} shows the details of the temperature and pressure properties of the EVMH. Both temperature and pressure are shown in simulation units ($\tilde{T_e} \rightarrow \dfrac{T_e\,k_B}{m_e\,c^2}$, and $\tilde{P_i} \rightarrow \dfrac{P_i\,1000}{m_i\,\omega_{pe}^3\,c}$, $\tilde{P_e} \rightarrow \dfrac{P_e\,1000}{m_e\,\omega_{pe}^3\,c}$, and magnetic pressure $P_B = \tilde{B}^2/8\pi$). Figure~\ref{fig:xsection2}(a) shows there is a sharp increase in electron temperature within the hole. The temperature cross section is not symmetrical, with an elevated electron temperature at $-400\lambda_D$, compared to $+400\lambda_D$ (see Fig.~\ref{fig:2Dproperties1}(b)). There are two peaks in the temperature profile (Fig.~\ref{fig:xsection2}(a)) which approximately match the location of the peaks in current at $\pm200\lambda_D$. Perpendicular electron temperature anisotropy, $T_{e\perp}/T_{e\|}$, (Fig.~\ref{fig:xsection2}(b)) increases within the EVMH from $\sim 0.7-0.9$ outside the structure (cf. Fig.~\ref{fig:Relative_track}) and increasing to $\sim 1.4$ inside the EVMH, remaining roughly constant between $\pm200 \lambda_D$ around its center.

Figure~\ref{fig:xsection2}(c) shows electron pressure, $P_e$ (red), ion pressure, $P_i$ (green), magnetic pressure, $P_B$ (blue) and total pressure, $P_t$ (black line). This figure shows that there is a reduction in total (plasma plus magnetic) pressure within the structure, which is mostly due to the reduction in magnetic pressure within the hole where the total field strength is reduced (Fig.~\ref{fig:xsection1}(a)). The magnetic pressure within the hole drops below both the electron and ion pressure. This results in a value of plasma beta, $\beta$, of $\sim5$ within the EVMH, whilst outside the hole plasma beta remains at $\sim1$.

Figure~\ref{fig:xsection2}(d) shows the off-diagonal terms of the electron pressure tensor. These terms appear to become important within the EVMH as $P_{xz}$ (green) takes on a large negative value, which is non-symmetric. Figure~\ref{fig:xsection2}(e) shows the main components of the pressure tensor, rotated into the local parallel (to \textbf{B}) and perpendicular directions. This confirms that there is an increase in perpendicular pressure within the structure, as both the $P_{\perp1}$ (green) and $P_{\perp2}$ (blue) lines are larger than the parallel pressure (red). The clear difference in the $P_{\perp1}$ and $P_{\perp2}$ values indicates non-gyrotropic behavior. The peaks in $P_{\perp1}$ correspond to the peaks in current in Fig.~\ref{fig:xsection1}, and therefore shows an increase of in-plane pressure due to the ring-shaped current.

This information indicates that the magnetic holes are being formed by a population of hot, possibly trapped electrons in the region, as there is little ion movement within the structure given the timescale of the simulation. The anti-clockwise electron drift (as indicated in Fig.~\ref{fig:2Dproperties1}(a)) would have a diamagnetic effect, and cancel out the magnetic field strength in the out-of-plane direction, hence the reduction in $B_z$.
\subsection{Electron velocity distributions and particle tracking.}\label{sec:turb_eVDF_tracking}
\begin{figure}[h!]
\includegraphics[scale=0.48]{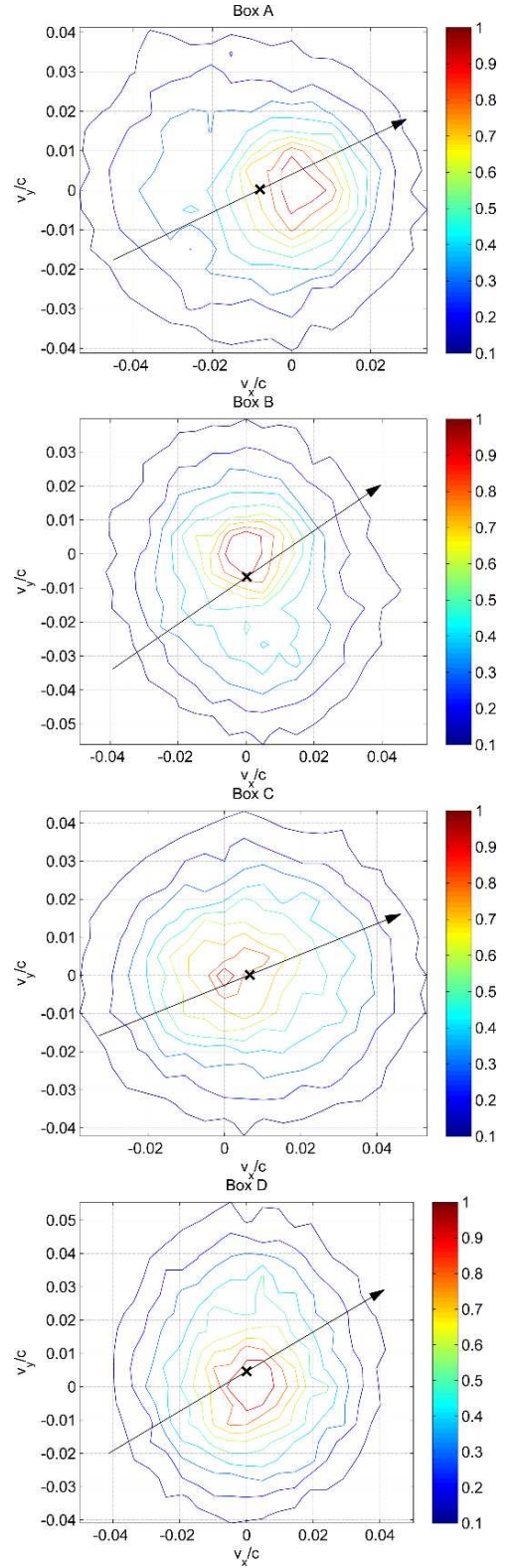}
\caption{\label{fig:VDF1} Electron velocity distribution functions in the $v_x$-$v_y$ plane for Boxes A, B, C and D. (Refer to Fig.~\ref{fig:2Dproperties1}(c).) The direction of the in-plane magnetic field is shown by a black arrow. The small black cross represents the electron bulk velocity.}
\end{figure}
Figure~\ref{fig:2Dproperties1}(c) shows the ring-like shape of the current formed within the EVMH. Four locations marked A, B, C, and D are selected around the structure in order to calculate electron velocity distribution functions (VDFs) at $t = 200\Omega_e^{-1}$, as shown in Fig.~\ref{fig:VDF1}. The maximum values of the VDFs have been normalized to 1 and the direction of the in-plane magnetic field is shown by a black arrow. The small black cross in each figure represents the electron bulk velocity in the $x$-$y$ plane. All four VDFs show a similar shape, but with different orientations. The black cross in Fig.~\ref{fig:VDF1} box A shows the electron bulk velocity is in the negative $x$ direction, in agreement with Fig.~\ref{fig:2Dproperties1}(a) which indicates anti-clockwise electron flows. Since box A is in a central, upper location of the vortex (Fig.~\ref{fig:2Dproperties1}(c)) the bulk velocity must be mainly in the negative $x$ direction for the flow to be anti-clockwise. This pattern holds for all four VDFs, with Fig.~\ref{fig:VDF1} box B indicating negative $y$ velocity, box C indicating positive $x$ velocity and box D indicating positive $y$ velocity.
\begin{figure}[h!]
\includegraphics[scale=0.6,angle=0]{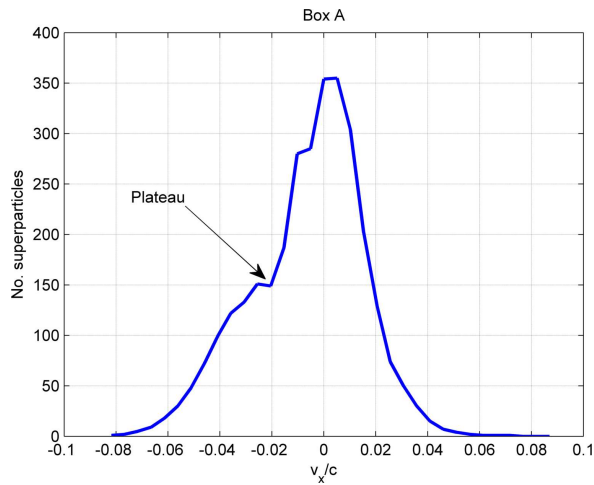}\\
\caption{\label{fig:xVDF} Maximum value of electron VDF along the $v_y$ direction, for Box A. (Refer to Fig.~\ref{fig:VDF1}.)}
\end{figure}

Each distribution shows a main peak, slightly offset in the opposite direction to the overall drift, and a plateau with extended tail of electrons in the appropriate anti-clockwise direction. To illustrate this, an effective cross section of the distribution in box A is shown in Fig.~\ref{fig:xVDF}. The data in this figure shows the maximum value of each column of data in the VDF for box A (Fig.~\ref{fig:VDF1}). Since this VDF is roughly symmetrical around $v_y=0$, it gives an impression of the cross section of this distribution along the $v_x$ direction. The main peak and plateau are clearly visible in Fig.~\ref{fig:xVDF}. The main peak appears to be approximately Maxwellian. The excess electrons populating the plateaus, in any of the four boxes, will have a higher perpendicular velocity than particles populating the core, as the guide field is approximately in the $z$ direction.

\begin{figure}
\includegraphics[scale=0.4,angle=0]{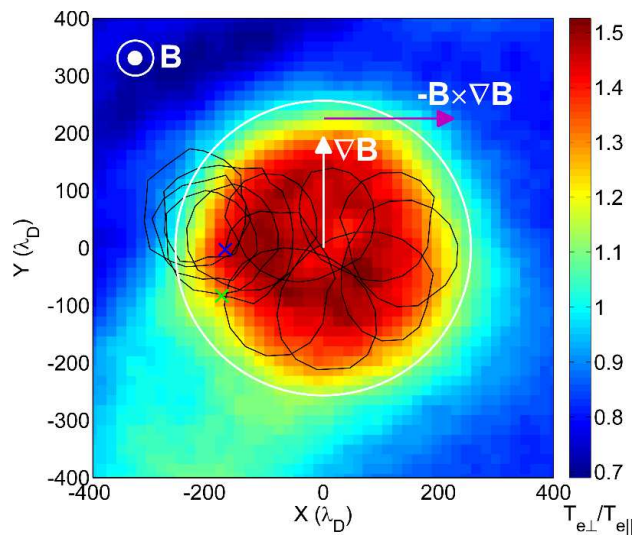}
\caption{\label{fig:Relative_track} An electrons trajectory (black) relative to the magnetic holes center. The green and blue crosses mark the start and end locations respectively. The circle (white) shows the approximate radius of the magnetic hole. The color scale shows $T_{e\perp}/T_{e\parallel}$ at time $t = 200\Omega_e^{-1}$.}
\end{figure}

In order to investigate the dynamics which have formed this structure, electrons were selected from the main peaks of all four boxes (A, B, C and D) and from a central location on each plateau, and their trajectories recorded from times $t = 100\Omega_e^{-1}$ to $t = 200\Omega_e^{-1}$. Animations of particles from the plateau portion of the VDFs suggests that some of these electrons are trapped, or partially trapped, within the magnetic hole. The EVMH is also slowly moving, from right to left, so its exact center was tracked by searching for the minimum value of $B_z$ in the region over the times of interest. Electron trajectories are plotted in a frame relative to the center of the magnetic hole, in order to look for trapped electrons. An example trajectory is shown in Fig.~\ref{fig:Relative_track} which shows an electron from the plateau region of the VDF from Box B (Fig.~\ref{fig:VDF1}).
 
In Fig.~\ref{fig:Relative_track} the black lines show the electrons relative trajectory. Note, that data was recorded at every 10th cycle, hence the jagged appearance in places. The white circle indicates the approximate size of the magnetic hole. The green cross shows the relative start location of the electron, and the blue cross shows its last recorded position. The background color shows a 2D map of electron perpendicular temperature anisotropy, $T_{e\perp}/T_{e\parallel}$, at $t = 200\Omega_e^{-1}$, showing that the highest perpendicular anisotropy corresponds to the region where the electron is trapped.

Figure~\ref{fig:Relative_track} shows that the electron initially follows a (relatively) stationary circular orbit, trapped along the left hand side of the hole, and then it eventually falls into a petal-like orbit with a guiding center motion clockwise relative to the vortex center. It is clear from this orbit, that only electrons with a gyroradius approaching that of the EVMH, can be influenced by it in this way. Electrons above a certain energy have the potential to enter these trapped orbits, as they sample reduced fields towards the holes center and increased field at its edges during a single gyro-orbit, the net effect being a drift perpendicular to the radial direction, and can be described as a cylindrically symmetric gradient drift. We have examined 676 electron trajectories from the plateaus of the VDFs of all four boxes in Fig.~\ref{fig:2Dproperties1}(c). Most of them show similar petal shaped orbits, but they are not always completely trapped. Some, like the electron in Fig.~\ref{fig:Relative_track} do not initially interact with the structure, but they eventually become trapped. Others initially interact, and can escape after time. Therefore they appear quasi-trapped. 

However, the drift of these trapped electrons is incompatible with the required direction of current, which is anti-clockwise, as shown in Fig.~\ref{fig:2Dproperties1}(a). In an EVMH field structure as the one seen here, $\nabla |\mathbf{B}|$ points in a direction from the center towards the edges of the EVMH. (See the vectors drawn in Fig.~\ref{fig:Relative_track}.) The direction of the gradient drift velocity, $\mathbf{v_{\nabla B}}$, is in the direction $\pm\mathbf{B} \times \nabla |\mathbf{B}|$, where the $\pm$ should be applied for the charge of the particle in motion. With the guide field out of the simulation plane, this results in a clockwise drift for an electron.

The total current is not due to the drift of any one electron, but it is due to the the collective motion of these trapped electrons, in combination with the gradient in $T_{e\perp}$ along the outer edge of the structure. We can infer the collective effect of many such trajectories by examining Fig.~\ref{fig:Relative_track} and consider that at the same moment in time, that there is an electron at every point of the plotted trajectory that will follow the same path. The trajectories along the radial direction often come close to electrons moving radially in the opposite direction, so their contribution to overall current will cancel. The contribution of trajectories along the outer edges of the ``petals'' are only reduced by non-trapped electrons outside the structure. This is because of the increased perpendicular anisotropy within the hole. Electrons with a center of gyration outside the hole have lower perpendicular energies, and their trajectories will not cancel all of this current at the edge of the ``petal''. This explains how a thin anti-clockwise ring-shaped current could be formed from this type of petal orbit. In order to create a much wider ring-shaped current, many different sized/shaped trapped orbits would be required within the MH at any one time.

An analysis was performed on a group of electrons from the plateaus of the VDFs (Fig.~\ref{fig:VDF1}) in order to confirm that these electrons are quasi-trapped within the MH. A total of 3076 electrons were used to generate images that revealed their distribution in space as the EVMH moved. Animations show that they have a circular distribution in space, matching the size and position of the EVMH, confirming they are (quasi) trapped. This evidence suggests that the quasi-trapped electrons are responsible for the ring shaped current in the EVMH. The hypothesis being that radially directed parts of the trajectories of the electrons within the hole are cancelled out by the presence of oppositely directed electron trajectories. Only azimuthal directions contribute to current at locations, such as at the edge of the petal orbits where no oppositely directed trapped orbits exists, and the non-trapped electrons outside the hole have lower energies so that the current at these locations cannot completely cancel. In order to show this a more idealized analysis of the types of trapped orbits is required.

\section{Test particle experiments}\label{sec:test_particle_sims}

In Section~\ref{sec:turb_eVDF_tracking} the orbits of the particles suggested that the collective effect of trapped or quasi-trapped electrons may be responsible for the ring-shaped current, as seen in Fig.~\ref{fig:2Dproperties1}(c). To test this hypotheses, a test particle code was developed that uses static magnetic and electric fields to calculate the trajectory of an individual particle, in order to categorize the various electron orbits near/within an EVMH.
\begin{figure}
\includegraphics[scale=0.31]{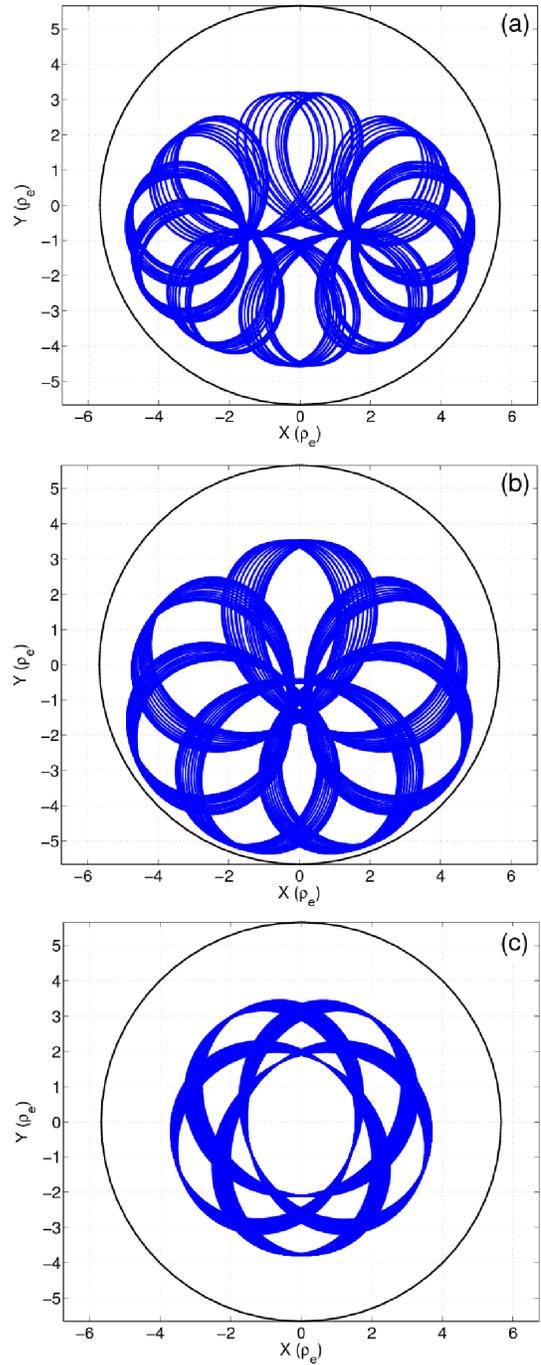}
\caption{\label{fig:TP_Trapped} Example test particle trajectories for electrons which are trapped by the EVMH.}
\end{figure}

A simple magnetic hole configuration was set up similar to those seen in the 2D turbulence simulations (Section~\ref{sec:turb_sims}). The magnetic field was directed in the $x$-$z$ plane, at an angle of \ang{15} to the $z$-axis, with a total magnetic field strength set to the arbitrary value of 1. Electric fields were set to zero. A circular magnetic hole was imposed in the simulation plane, with a quarter-period sinusoidal radial profile and a $50\%$ drop in $B_z$ at its center, compared to $B_z$ outside the hole, consistent with Fig.~\ref{fig:xsection1}. The radius of the hole, $R_H$, was set to match that seen in the turbulence simulation, which was $\sim300\lambda_D$, or $\sim 6 \rho_e$. The electron thermal Larmor radius, $\rho_e$, was determined in the test code by setting the value of electron beta, $\beta_e$, equal to $0.5$ and assuming isotropic temperature. Velocities are normalised to the speed of light, $c$, and time is normalised to the electron plasma frequency, $\omega_{pe}$. The electrons were initially positioned randomly within a radius twice that of the EVMH, so that the behavior of electrons entering or escaping the MH and those trapped inside could be investigated. Initial velocities were also randomised, using an isotropic Maxwellian distribution function so that statistical data could be gathered. Electron trajectories were calculated and recorded for a maximum of $200$ gyroperiods ($1256\Omega_e^{-1} $), or until the electron was $\sim 7R_H$ away from the center of the hole, in order to reduce unnecessary calculation time for electrons that are not trapped and escape the hole. The timestep, $\Delta t$, used to calculate each particle's position and velocity was chosen such that $200$ steps were calculated for every gyroperiod.

This analysis showed that there are several types of electron trajectory possible near a MH. The first are trajectories that are not trapped by the MH. An electron's perpendicular velocity,  $v_\perp$, initial position and direction are all important factors determining if it will become trapped. When an electron's $v_\perp$ is small, such that its gyroradius is considerably smaller than the MH radius, $R_H$, then the MH only slightly deflects the electrons path and it does not become trapped. If an electrons position is outside the hole and its initial velocity is such that it gyrates away from the hole, then it will also not become trapped.

We also found example electron trajectories that are quasi-trapped by the MH. These can be split into two types, those whose orbits are not entirely contained within the MH, but do not escape and electrons that are significantly diverted/reflected by the MH but eventually escape. These electrons have larger $v_\perp$ than electrons classed as not trapped. This supports the theory suggested in Section~\ref{sec:turb_eVDF_tracking} that electrons with Larmor radii approaching that of the hole radius are more likely to become trapped by it.

The final class of electrons are those that that are fully trapped within the MH, as shown in Fig.~\ref{fig:TP_Trapped}. These have similar $v_\perp$ to the quasi-trapped electrons, but have initial positions and directions, so that the electrons can form ``petal'' shaped orbits. The trajectories in Fig.~\ref{fig:TP_Trapped}(a) and (b) most closely resemble the example trajectory seen in the turbulence simulation in Fig.~\ref{fig:Relative_track}. The relative magnitudes of the electron perpendicular velocity, $v_\perp$, can be inferred from the radii of their gyration. It is clear to see that the electron in Fig.~\ref{fig:TP_Trapped}(a) has a smaller $v_\perp$ than that in Fig.~\ref{fig:TP_Trapped}(b), and it forms a trajectory with more petals, and takes a larger number of gyroperiods to process completely around the MH. Figure~\ref{fig:TP_Trapped}(c) shows that initial position and velocity is also a factor, as given the correct conditions, highly symmetrical trajectories can be produced. Our analysis also shows that completely circular trajectories without forming petal-shaped orbits are possible.
\begin{figure}
\includegraphics[scale=0.6]{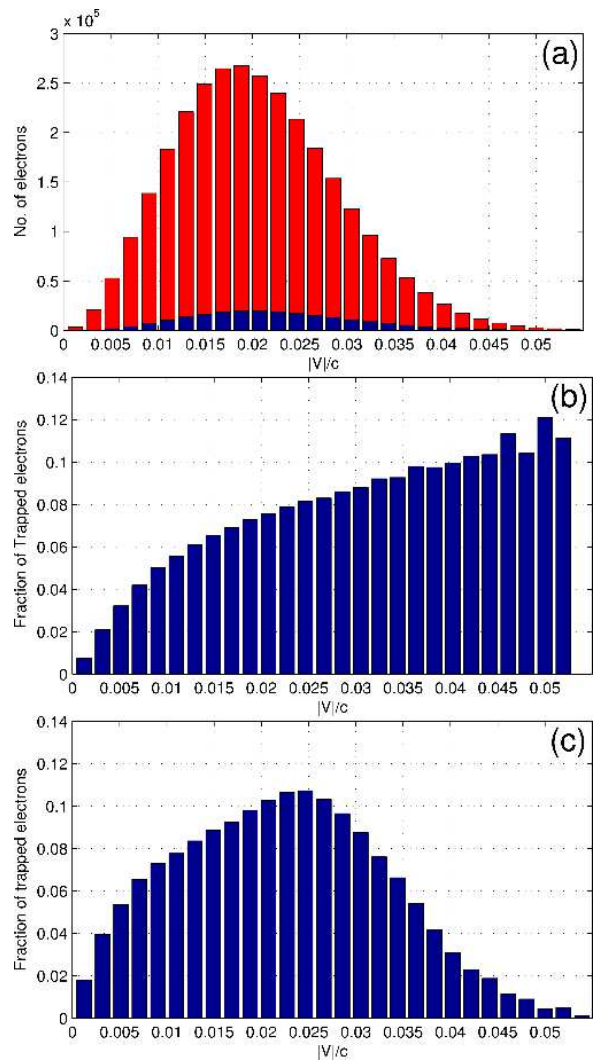}
\caption{\label{fig:Mag_V_distributions}(a) Speed distribution of total (red) and trapped (blue) electrons for a hole with radius, $R_H \sim 6 \rho_e$. (b) Distribution of the fraction of trapped electrons for a hole with radius, $R_H \sim 6 \rho_e$. (c) Distribution of the fraction of trapped electrons for a hole with radius, $R_H \sim 3 \rho_e$.   Velocities are normalised to the speed of light, c.}
\end{figure}

Whilst examining trajectories can help us to understand the basic electron dynamics and individual behavior within an EVMH, we also used a statistical approach in order to better characterize the population of trapped electrons in these structures. A test was developed that determined automatically whether a particle was part of the trapped population. This was done according to the following rules: If the electrons final position is within a square box of length four times the EVMH radius (centered on the hole) after 200 gyroperiods, then the particle is considered trapped. This ensures that quasi-trapped particles that can orbit considerably outside the hole are considered part of the population, whilst those that eventually escape are discounted. Additional checks also ensure that electrons with extremely small gyroradii are not classed as trapped as they do not contribute much to the current within the MH. 

In order to build up statistics that are comparable to the PIC studies (Section~\ref{sec:turb_sims}), we use a similar number of test particles as in the previous simulation: a hole radius of 17 cells with 6400 particles per cell, results in $\sim 6\times10^6$ super-particles in the EVMH. Our initial work with the test code showed that $\sim10\%$ of test particles become trapped, so to be able to demonstrate the shape of any instantaneous current due to trapped electrons, we would like to have trapped trajectory data for $\sim \num{6e5}$ electrons.

The statistical data we show used an initial population of $\num{3e6}$ electrons, initially distributed randomly within a radius of $2R_H$, with velocities set to an isotropic Maxwellian distribution. This results in a speed distribution that ranges from $\sim 0-4.5 v_{th}$. Using the selection method described above, the code recorded $\sim \num{2e5}$  trapped electrons, which is of the order of the number required for comparison with the turbulence simulation. The speed distribution of the total number of electrons (red), and trapped population (blue) is shown in Fig.~\ref{fig:Mag_V_distributions}(a). This graph confirms that the initial population was in a Maxwellian distribution.

Plotting the fraction of trapped electrons (Fig.~\ref{fig:Mag_V_distributions}(b)) confirms the earlier statements that the larger the velocity, the more likely the particle will be trapped in the EVMH, as the histogram follows a smoothly increasing curve. The maximum fraction of trapped electrons is $12\%$, with an average of $7.3\%$. Figure \ref{fig:Mag_V_distributions}(c) shows how the trapped population changes for a smaller sized MH with radius $R_H \sim 3 \rho_e$. This shows that an EVMH is selective, and has a preferred energy range for electrons it can trap. This makes physical sense, as particles with larmor radii approaching or greater than the hole radius should only be deflected by the hole rather than enter into trapped, petal-like orbits. This trend is not visible in Fig.~\ref{fig:Mag_V_distributions}(b) because of the lack of test particles at higher energies, as set by the Maxwellian distribution.  

Particles with $|v|/c > 0.055$ are not shown in Fig.~\ref{fig:Mag_V_distributions} as the fraction becomes statistically unreliable, producing erroneous data spikes. The total number of electrons in each bin shown in Fig.~\ref{fig:Mag_V_distributions}(a) is greater than 1000.
\begin{figure}[h!]
\includegraphics[scale=0.6,angle=0]{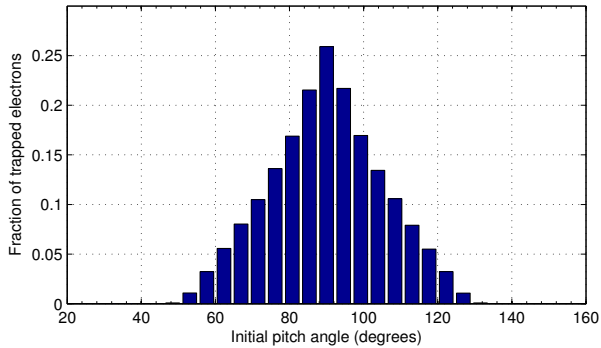}
\caption{\label{fig:Trapped_pitch} Distribution of initial pitch angles for trapped electrons.}
\end{figure}

We now examine the initial pitch angle, defined as $\arctan(v_\perp/v_\parallel)$, of the trapped electron population. The Maxwellian distribution algorithm produces an initial distribution of pitch angles peaking at \ang{90} in a sinusoidal form. Figure~\ref{fig:Trapped_pitch} shows the ratio of the number of trapped particles to the number of total particles in each angular bin. This shows a distinctive peak, that is symmetric around \ang{90}, meaning that particles with a large ratio of $v_\perp$ to $v_\parallel$ are more likely to get trapped. Therefore, within the EVMH, there will be an increase in the number of electrons that have a high value of $v_\perp$. This, as we have seen in Fig.~\ref{fig:xsection2}, will result in an increase in perpendicular pressure, and therefore $T_{e\perp}$ within the hole. This also explains why a high value of $T_{e\perp}/T_{e\parallel}$ was observed within these EVMH structures in Figs.~\ref{fig:xsection2} and \ref{fig:Relative_track}.
\begin{figure}
\includegraphics[scale=0.45,angle=0]{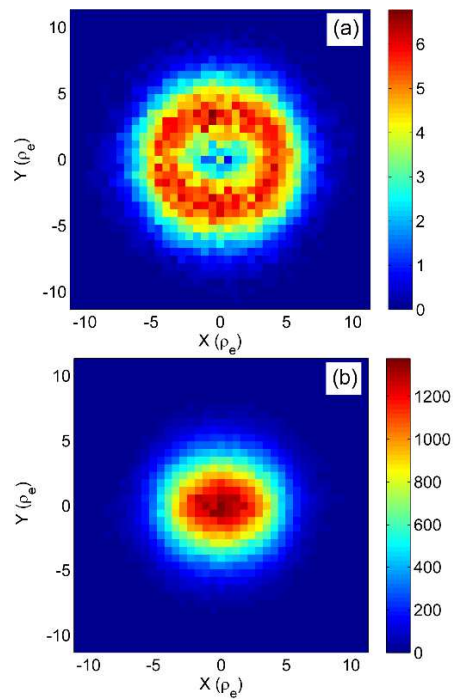}
\caption{\label{fig:Current_sum_3000}(a) Total current calculated from the trapped electron trajectories. (b) The number of electrons in each bin, at $t=94\Omega_e^{-1}$.}
\end{figure}

To see if the collective trajectories of these trapped particles could be responsible for the ring-shaped current, a series of 2D histograms was created using the test particle code. A square box, $4R_H$ wide was centered around the EVMH and divided into $40\times40$ square cells. After the position of a test particle was calculated, its location within this grid was determined and various velocity sums calculated for each grid point, (e.g. $J_x \sim \sum v_x = N\langle v_x \rangle$) in order to build up a picture of the instantaneous current these trapped electrons could create.

Figure~\ref{fig:Current_sum_3000}(a) shows the magnitude of the total current calculated from the population of trapped electrons binned into a 2D histogram, at $t=94\Omega_e^{-1}$. This shows that the combination of multiple trapped orbits sum to form a distinctive ring current, mostly contained within the radius of the magnetic hole ($\sim 6\rho_e$). This matches the form of current seen in the turbulence simulation in Fig.~\ref{fig:2Dproperties1}(c). A similar histogram (not shown) that sums only the azimuthal component of the electrons in-plane velocity, also forms a similar ring shape. A histogram (not shown) that sums only the radial component of velocity, does not form a ring-shape, but instead is circular, with values approximately 1/10th that of the total current. This confirms the hypothesis made at the end of Section~\ref{sec:turb_eVDF_tracking} that the radial parts of the electron trajectories within the MH cancel out when the average current is calculated, and it is the azimuthal component that is the strongest contributor to the overall current. 
\begin{figure}[h!]
\includegraphics[scale=0.4,angle=0]{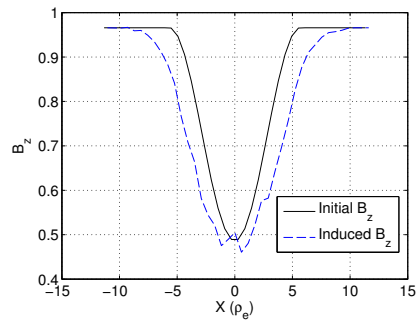}
\caption{\label{fig:Bz_xSection} Initial and induced $B_z$ cross section.}
\end{figure}

The out-of-plane magnetic field that would be induced by this current was calculated using the equation $\mathbf{J}\sim \nabla\times \mathbf{B}$, and the in-plane components of current shown in Fig.~\ref{fig:Current_sum_3000}, using an integration method. The ring-shaped current creates a circular depression in the $B_z$ component of the magnetic field (not shown). A cross section of the initial $B_z$ profile and the induced $B_z$ is shown in Fig.~\ref{fig:Bz_xSection}. A weighting of \num{0.0032} was applied to each test particle in order to match the depth of the $B_z$ depletion.

Due to the fact that test particles can travel outside the hole radius and still remain trapped, the induced $B_z$ profile is slightly wider than the initial hole width. This is only an estimate of what fields might be produced in a self-consistent model. There will be electrons, not trapped, gyrating outside the hole (not included in this calculation of current) that form opposing currents at the outer edges of these trapped orbits, which would alter (perhaps reduce further) the amount of $B_z$ depletion at the edge of the structure. However, the similarity of the original imposed field profile and that inferred from the trapped particles suggests that it might be possible for electrons trapped in such a structure to find a stable (or quasi-stable) Vlasov equilibrium.

Figure~\ref{fig:Current_sum_3000}(b) shows a similar 2D histogram as Fig.~\ref{fig:Current_sum_3000}(a) but sums the electron counts in each bin location at the same timestep. This value will be proportional to number density multiplied by a weighting factor. This shows a high density of trapped electrons at the center of the EVMH, gradually reducing with circular symmetry towards the outer edges of the hole. This matches the previous observation in the turbulence simulation of increased electron density in the EVMH center in Fig.~\ref{fig:xsection1}(c). The increased electron density in the region suggests that there should also be an increased ion density in the EVMH in order to maintain charge neutrality, for $\mathbf{E}=0$, but the electron test particle analysis does not provide any information about this. The data in the turbulence simulation in which these were first observed, suggests that there was a corresponding ion density increase in the area, but this is of the order of the other random density fluctuations in the simulation. It is difficult to say whether this increased ion density is a crucial factor in the formation of an EVMH. To investigate if an EVMH can be formed, and reach a natural Vlasov equilibrium, without the presence of turbulence, a series of self consistent simulations were performed.
\section{Simulations of Isolated Structures}
\label{sec:isolated_sims}
We now show results from PIC simulations, using Parsek2D \cite{markidis:2009}, designed to recreate a simple EVMH in a quiet uniform plasma. Using PICs code in this way is a useful method of finding a Vlasov equilibrium when analytical solutions are hard to derive. The intent is to show that an EVMH can be ``seeded'' within the plasma by adjusting the initial properties of the plasma in a localised area. The simulations were initialized with the same plasma properties as in the initial turbulence simulation (Section~\ref{sec:turb_sims}), but in a smaller box with a size of $\sim$ 1/2 ion inertial length, and an increased number of particles per cell ($10000$).
\begin{figure}
\includegraphics[scale=0.3]{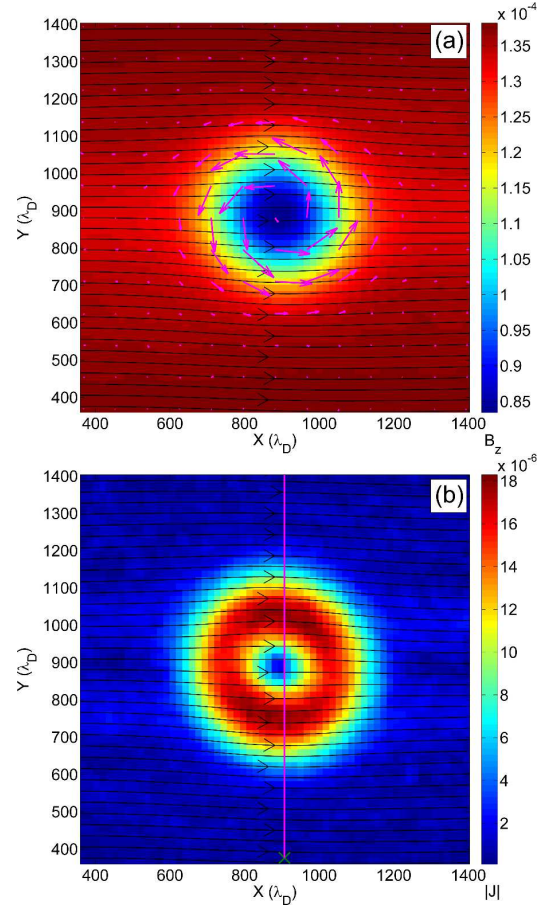}
\caption{\label{fig:2Dproperties_toy} Magnetic field lines (black) and (a) out-of-plane magnetic field $B_z$ with electron flow vectors (magenta), (b) magnitude of current $J$ at $t=250\Omega_e^{-1}$. The vertical line shows where cross sections of the parameters will be taken.}
\end{figure}

A number of PIC simulations were run, designed to examine what plasma properties are required in order for an EVMH to achieve an equilibrium. These tests showed that a depletion in magnetic field strength alone was not enough to seed a stable EVMH. A stable vortex also required an increase in perpendicular electron temperature anisotropy within the same region.

In the simulation results shown here, the magnetic field is configured with similar properties used in the Section~ \ref{sec:test_particle_sims}. The domain has a quiet magnetic field set at \ang{15} to the $z$ axis in the $x$-$z$ plane. This was seeded with a circular, half sinusoidal drop in the $B_z$ component of magnetic field, with a maximum drop at its center of 50\%. The ions were initialized with isotropic Maxwellian velocity distributions. The thermal velocities of the electrons were adjusted to produce a perpendicular temperature anisotropy of $0.8$ outside the hole, and $1.4$ inside the hole. These parameters were successful at forming a stable MH, which remains stationary at the center of the simulation.

Figure~\ref{fig:2Dproperties_toy} shows the simulation results at time $t=250\Omega_e^{-1}$. Panel~(a) shows the circular depletion in $B_z$ and Panel~(b) shows the current density within the vortex, which again has a characteristic ring shape. The vertical line, as in Section~\ref{sec:Turb_params}, shows the location where a cross section of plasma parameters will be taken. Overall, the final configuration is remarkably similar to that seen in the turbulence simulations (Fig.~\ref{fig:2Dproperties1}).
\begin{figure}[t]
\includegraphics[scale=0.6,angle=0]{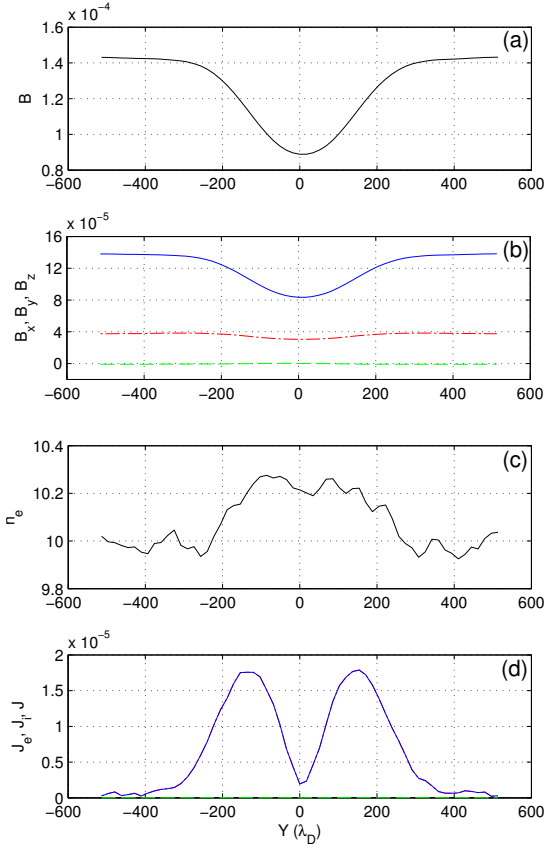}
\caption{\label{fig:xsection1_toy} Cross section of field and current parameters through the line indicated on Fig.~\ref{fig:2Dproperties_toy}(b).}
\end{figure}
\begin{figure}[t]
\includegraphics[scale=0.6,angle=0]{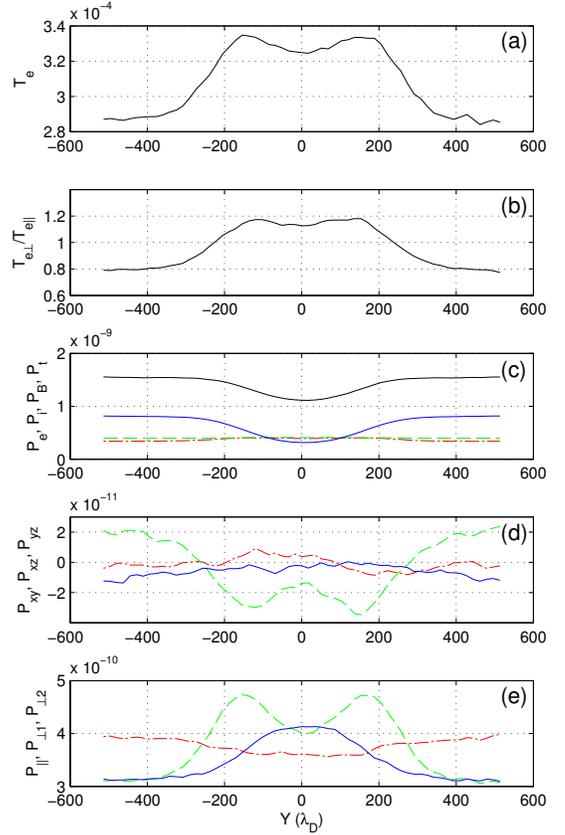}
\caption{\label{fig:xsection2_toy} Cross section of plasma temperature and pressure parameters through the line indicated on Fig.~\ref{fig:2Dproperties_toy}(b).}
\end{figure}

Figures~\ref{fig:xsection1_toy} and \ref{fig:xsection2_toy} show cross sections of the properties of the EVMH through the line shown in Fig.~\ref{fig:2Dproperties_toy}(c), in the same style as Figs.~\ref{fig:xsection1} and \ref{fig:xsection2}. These show a 40\% reduction in $B_z$ (Fig.~\ref{fig:xsection1_toy}(b)) and a $\sim3\%$ increase in electron density within the structure (Fig.~\ref{fig:xsection1_toy}(c)). The current is again formed by electron flow, as the $J$ (blue) and $J_e$ (red) lines completely overlap (Fig.~\ref{fig:xsection1_toy}(d)).

Figure~\ref{fig:xsection2_toy}(a) shows the increased electron temperature within the hole, and Fig.~\ref{fig:xsection2_toy}(b) shows the increased perpendicular temperature anisotropy. The magnetic pressure reduces below the electron and ion pressure (Fig.~\ref{fig:xsection2_toy}(c)) as seen in Section~\ref{sec:Turb_params}. Nonsymmetric off-diagonal terms are again present in the pressure tensor (Fig.~\ref{fig:xsection2_toy}(d)), and the difference in the $P_{\perp1}$ and $P_{\perp2}$ components of the pressure tensor show non-gyrotropic behavior.

An increase in ion density (not shown) matching that of the electrons is also present in the MH, which implies that an electrostatic potential has been set up in the region, in order to reach a stable equilibrium. In order to see if this was the case, the in-plane components of electric field, $E_x$ and $E_y$ were integrated in order to calculate the electrostatic potential, $\Phi$, using the equation $\mathbf{E}=-\nabla \Phi$. This showed that a negative potential has been created within this EVMH. This means that positively charge particles would actually lose potential energy as they enter this region, and are attracted/accelerated into the area. This explains the increase in ion density in the region, which matches the electron density in the hole. No similar potential was observed in the turbulence simulations (Section~\ref{sec:turb_sims}) which suggests that this is not a requirement for EVMH formation, and is likely a result of not achieving the exact same solution of the Vlasov equation, as seen within turbulence.

Calculating the electron VDFs reveals similar distributions to those seen in Section~\ref{sec:turb_eVDF_tracking}. Particle tracking also shows a range of trapped trajectories, of both the petal and circular type. Electrons with higher perpendicular velocities tend to circular, whereas lower energy electrons follow petal shaped orbits, as discussed in Section~\ref{sec:test_particle_sims}. 

These simulations show that a circular sinusoidal magnetic depletion in $B_z$ can form an electron structure, in Vlasov equilibrium that is stable over many electron gyroperiods. This structure will only be stable if it is populated by electrons with a higher perpendicular temperature anisotropy that those outside the EVMH. Previous simulations have demonstrated that electron temperature anisotropy can fluctuate within turbulence, and that sub-ion-scale reconnection can produce localised regions of high parallel temperature anisotropy\cite{Haynes:2014}. Therefore EVMHs can naturally form within plasma where magnetic field strength is depleted in areas of high $T_{e\parallel}/T_{e\perp}$.

\section{Conclusions}
We have presented the results of 2-D simulations that show coherent nonlinear magnetic structures can form within sub-proton-scale plasma turbulence with a guide field. The structures are circular, have a depletion in $B$ in the guide field direction, and contain a population of hot electrons with a characteristic ring shaped current density. These ``magnetic holes" are $\sim300 \lambda_D$ or $\sim6 \rho_e$ in radius. The current density is entirely formed by mean azimuthal electron flow, and thus we call them electron vortex magnetic holes (EVMHs). 

The EVMH structures have a high perpendicular electron temperature anisotropy within them, and an electron density increase of $\sim 10\%$. There is a dip in total pressure, $P_t$ in the region, resulting in increased values of $\beta$ within the hole. Magnetic pressure is seen to dip below both the ion and electron pressure. Off-diagonal terms of the pressure tensor are non-zero, and non-symmetric, and the pressure tensor rotated into the parallel and perpendicular directions shows non-gyrotropic behavior. Particle VDFs show a Maxwellian-like part of the distribution and an additional population of high $v_\perp$ current-carrying electrons, forming a plateau and extended tail in the distribution function. Tracking particles confirms that these electrons are trapped within the hole, and trajectory plots show that they follow petal-shaped orbits. These orbits are the result of the reduced $B_z$ field in the region, and are similar to orbits produced by electrons drifting in a magnetic field gradient, where the field gradient here is circularly symmetric. An analysis of groups of particles from various parts of the VDFs confirmed that the higher-energy electrons were a group of trapped electrons, potentially responsible for the circular ring current in the region, as they collectively followed the circular form of the vortex as the MH drifted in space. 

Test particle simulations were used to demonstrate the different types of trapped trajectories which are possible within a static, circular sinusoidal depletion in magnetic field. For an ensemble of test particles one can calculate the effective current and density that they would produce (assuming some particle weighting). Assuming the test particles have a Maxwellian velocity distribution at some temperature, each type of orbit contributes to the instantaneous current with the EVMH. We showed that these collectively form a ring-shaped current within the MH. The ring-shaped current has a diamagnetic effect, and can induce a similarly shaped magnetic field to the initial drop in $B_z$, implying a solution to the Vlasov equation might be possible in a self-consistent field model.
Using a statistically significant number of particles, we showed that $\sim7\%$ of test particles become trapped, and that the probability of a particle being trapped was increasing with the magnitude of its velocity. We also showed that EVMHs selectively trap electrons of a certain energy range, as the trapped electron distribution develops a peak for smaller hole radii. We also found that particles with \ang{90} pitch angles are most likely to be trapped within these structures, and a small increase in the electron density should be observed within the structure due to the overlapping orbits of the trapped electron population.

We then showed that we can seed a stable EVMH in a self-consistent PIC simulation with a ``quiet'' plasma background, using a perturbation consisting of a circular drop in $B_z$ and increase in perpendicular electron temperature anisotropy within the hole. These results show that $B_z$ induced by these trapped electrons appears to reach a quasi-stable solution, at least on sub-proton timescales, so that an EVMH can exist without the presence of turbulence. The trapped population of electrons within the MH generates a negative electrostatic potential, which attract ions into the region, allowing an increase in electron density and corresponding ion density increase. However, no negative potentials were seen in the holes within the turbulence simulations which suggests that this potential is not a necessary requirement for the stability of an EVMH. 

A recent survey in the Earth's plasma sheet by \citet{Sun:2012} showed that sub-proton scale MHs were common occurrences. The observed population had a range of sizes, with a peak in the frequency distribution at $0.5\rho_i$, where $\rho_i$ is the ion thermal larmor radius. These events also showed an increase in electron energy flux at a pitch angle of $\ang{90}$ inside the MHs, indicating that electron dynamics could play an important role in their formation. The simulated EVMHs in this paper have a diameter of $\sim 0.3 \rho_i$ and are therefore consistent with their findings. The increase in electron flux is also consistent with the findings of Section~\ref{sec:test_particle_sims} that \ang{90} pitch angle particles are more likely to be trapped. A detailed multi-spacecraft analysis by \citet{Torbjorn:NEW} of sub-proton-scale MHs in the plasma sheet shows that the \citet{Sun:2012} observations closely match the predictions of EVMHs as described in this paper, whilst ruling out many other formation theories. In future work we will investigate what factors determine the radius of an EVMH, and how large they can be. 

We conclude that the EVMHs described in this paper provide a theoretical model for the sub-proton-scale MHs seen in the terrestrial plasma sheet, and may also be applicable to sub-proton-scale magnetic holes observed elsewhere in the heliosphere. These coherent structures may also affect turbulence statistics, such as intermittency, at electron scales and could be relevant for many types of astrophysical plasmas.

\begin{acknowledgments}
Some of the work reported here was carried out whilst CTH was supported by an STFC (UK) studentship. DB was supported by partial funding from STFC grant ST/J001546/1.
The research presented here has received funding from the European Commission's Seventh Framework Programme FP7 under the grant agreement SHOCK (project number 284515).
\end{acknowledgments}

\newcommand{\apjl}{{Astrophys. J. Lett.}}
\newcommand{\apj}{{Astrophys. J.}}
\newcommand{\grl}{{Geophys. Res. Lett.}}
\newcommand{\solphys}{{Solar Phys.}}
\newcommand{\jgr}{J.\ Geophys.\ Res.}
\newcommand{\mnras}{MNRAS}
\newcommand{\nat}{Nature}

 \newcommand{\noop}[1]{}
%


\end{document}